\documentclass[twoside]{article}

\usepackage[accepted]{aistats2025}

\usepackage[utf8]{inputenc} % allow utf-8 input
\usepackage[T1]{fontenc}    % use 8-bit T1 fonts
\usepackage{hyperref}       % hyperlinks
\usepackage{url}            % simple URL typesetting
\usepackage{booktabs}       % professional-quality tables
\usepackage{amsfonts}       % blackboard math symbols
\usepackage{nicefrac}       % compact symbols for 1/2, etc.
\usepackage{microtype}      % microtypography
\usepackage{xcolor}         % colors

\usepackage{amssymb}
\usepackage{amsfonts}
\usepackage{amsmath}
\usepackage{amsthm}
\usepackage{graphicx}
\usepackage{arydshln}
\usepackage{tabularray}
\usepackage{MnSymbol}
\usepackage{wasysym}
\newcommand{\KL}[0]{\operatorname{KL}}
\newcommand{\ent}[0]{\mathrm{H}}
\usepackage{amsthm}
\newtheorem{definition}{Definition}
\newtheorem{proposition}{Proposition}
\newtheorem{lemma}{Lemma}

\newcommand\independent{\protect\mathpalette{\protect\independenT}{\perp}}
\def\independenT#1#2{\mathrel{\rlap{$#1#2$}\mkern2mu{#1#2}}}
\begin{document}

\runningauthor{Zhaolu Liu, Mauricio Barahona, Robert L.\ Peach}

\twocolumn[
\aistatstitle{Information-Theoretic Measures on Lattices for Higher-Order Interactions}
\aistatsauthor{ $\text{Zhaolu Liu}^1$ \And  $\text{Mauricio Barahona}^1$ \And $\text{Robert L. Peach}^{2,3\, \ast}$}
\aistatsaddress{\And  $^1$Department of Mathematics, Imperial College London, United Kingdom \And}
\vspace{-2.1em}
\aistatsaddress{\And  $^2$Department of Neurology, University Hospital W\"urzburg, Germany \And}
\vspace{-2.1em}
\aistatsaddress{\And  $^3$Department of Brain Sciences, Imperial College London, United Kingdom \And}
\vspace{-2.1em}
\aistatsaddress{\And  $^\ast$Corresponding author: peach\_r@ukw.de \And}
]

\begin{abstract}
Traditional measures based solely on pairwise associations often fail to capture the complex statistical structure of multivariate data.
Existing approaches for identifying information shared among $d>3$ variables are frequently computationally intractable, asymmetric with respect to a target variable, or unable to account for all the ways in which the joint probability distribution can be factorised.
Here we present a systematic framework based on lattice theory to derive higher-order information-theoretic measures for multivariate data. Our construction uses lattice and operator function pairs, whereby an operator function is applied over a lattice that represents the algebraic relationships among variables. We show that many commonly used measures can be derived within this framework, yet they fail to capture all interactions for $d>3$, either because they are defined on restricted sublattices, or because  
the use of the KL divergence as an operator function, a typical choice, leads to undesired disregard of groups of interactions. 
To fully characterise all interactions among $d$ variables,
we introduce the Streitberg Information, which is defined over the full partition lattice and uses generalised divergences (beyond KL) as operator functions. 
We validate the Streitberg Information on synthetic data, and illustrate its application in detecting complex interactions among stocks, decoding neural signals, and performing feature selection in machine learning.

\end{abstract}

\section{Introduction}\label{sec: intro}
Understanding the emergent structure of complex systems is a key objective across scientific disciplines, and traditional models built on pairwise interactions have been instrumental in revealing fundamental system effects~\cite{newman2011structure}. However, recent evidence suggests that many real-world systems cannot be adequately described by pairwise interactions alone~\cite{battiston2020networks,rosas2022disentangling}, much like a symphony requires the harmonisation of an entire orchestra, not just two musicians. In biology~\cite{schneidman2006weak}, ecosystems~\cite{grilli2017higher}, and social networks~\cite{benson2016higher}, higher-order interactions (i.e., those involving interactions between groups of more than two elements) are increasingly recognised as critical for driving collective dynamics~\cite{battiston2020networks,rosas2022disentangling,grilli2017higher,bick2023higher}. Yet direct measurements of such group interactions 
are rarely available, 
necessitating approaches to reconstruct the underlying interactions and mechanisms from observed multivariate data~\cite{rosas2022disentangling, santoro2023higher, young2021hypergraph, wang2022full}.

Information theory provides a rigorous mathematical framework to detect statistical interactions between variables with the  added capacity to relate statistical structure to function~\cite{timme2014synergy}. 
Early information-based metrics, such as \textit{total correlation} (TC)~\cite{watanabe1960information} and
\textit{interaction information} (II)~\cite{mcgill1954multivariate}, began to address higher-order interactions inspired by the notion of statistical independence, but their formulations considered only particular factorisations of the joint probability distribution (see Section~\ref{sec: lattice theory}), hence with restricted ability to capture complex interactions between multiple variables~\cite{krippendorff2009information, james2017multivariate,liu2023interaction}. These measures were later followed by \textit{connected information} (CI), which captured interactions among $d$ variables by considering their maximum entropy states~\cite{schneidman2003network}. Yet its practical application to real-world data is hampered by the computational difficulty of constructing the maximum entropy distributions~\cite{schneidman2003synergy}.
One of the prevalent measures is \textit{ partial information decomposition} (PID), which offers a breakdown of higher-order interaction into information atoms~\cite{williams2010nonnegative}. 
While PID details how information is shared among variables, it becomes computationally infeasible for $d>3$ and exhibits asymmetry among variables. 
The increasing interest in detecting higher-order interactions in data and the limitations of existing tools thus underscore the need for practical information-theoretic measures that can handle systematically such interactions for multivariate data by considering all factorisations of the joint probability distribution. 

Our contributions in this work are twofold: (i) First, we develop a mathematical framework that systematically formalises higher-order information-theoretic measures using lattice theory. 
We demonstrate that many pre-existing information-theoretic measures can be derived from sublattices of the partition lattice with divergences as operator functions over these sublattices.
(ii) Second, we propose the Streitberg Information (Equation~\eqref{eqn: SI}), a novel information measure based on the complete partition lattice, 
which accounts for all factorisations of the joint probability distribution of $d$ variables and vanishes when any lower-order factorisation is present.
We further demonstrate that, if the Kullback-Leibler (KL) divergence is used as the operator function (a typical choice), it leads to the collapse of lattice substructures, resulting in the undesired, non-trivial cancellation of groups of interactions. 
Therefore, to avoid this loss of information and to preserve the structure of the full partition lattice, the \textit{Streitberg Information} (SI) uses the Tsallis-Alpha divergence and we develop its consistent estimator using a $k$-nearest-neighbours method. Finally, we present numerical validation of SI on synthetic data, followed by illustrative applications to finance, neuroscience, and machine learning.

\section{Information-Theoretic Measures}\label{sec: 2}
In information theory,
the shared information between two random variables, 
$X_1$ and $X_2$,
can be quantified using the mutual information (MI):
\begin{align*}
    \mathrm{MI}(X_1;X_2) = \ent(X_1) + \ent(X_2) - \ent(X_1, X_2),
\end{align*}
where $\ent(\cdot)$ is the entropy of the random variables.  Mutual information vanishes if and only if the two variables are independent~\cite{berrett2019nonparametric}.

Generalising mutual information to capture dependencies among a set $\mathbf{\mathcal{X}}=\{X_1,\ldots, X_d\}$ of $d>2$ random variables is nontrivial.
A straightforward way to quantify the total shared information among $d$ variables is through total correlation (TC)~\cite{watanabe1960information}:
\begin{equation}\label{eqn: TC}
    \mathrm{TC}(d)=\sum_{i=1}^d \ent(X_i) - \ent(X_1, \ldots, X_d), 
\end{equation}
which vanishes if and only if the variables are jointly independent.
However, $\mathrm{TC}(d)$  
provides limited insight into the specific nature of the interactions that may exist between variables
~\cite{schneidman2003network, garner1962uncertainty}.
An alternative extension of MI is the interaction information (II)~\cite{mcgill1954multivariate}, which captures the higher-order information in a system of $d$ variables through the inclusion-exclusion principle:
\begin{align}
    \mathrm{II}(d)=-\sum_{T \subseteq \mathcal{X}}(-1)^{d -|T|} \, \ent(T),
    \label{eq:II}
\end{align}
where the sum extends over the power set of all subsets $T$ of $\mathcal{X}$ and $|T|$ denotes the cardinality of set $T$.
Although widely used, the interpretation of the vanishing of $\mathrm{II}(d)$ is not clear~\cite{krippendorff2009information}. Indeed, we show below that this is because non-trivial higher-order terms are decomposed into a sum of lower-order terms via KL divergence (Section~\ref{sec: 4}).
We also show below that, although seemingly distinct, both TC and II actually belong to the same family of information-theoretic measures when formalised through lattice theory (Section~\ref{sec: lattice theory}). 

\section{Lattice Theory}\label{sec: lattice theory}
To understand how different information-theoretic measures
%(e.g.,those in Section~\ref{sec: 2})
can be systematically generated from different lattices, we first provide a few relevant concepts and results from lattice theory. 

A partially ordered set (poset) defined on $\mathcal{X}$ with partial order $\leq$ is a lattice (denoted by $\mathcal{L}$) if 
any two elements $x,y\in \mathcal{L}$ have a unique supremum $x\vee y$ (called join) and a unique infimum $x\wedge y$ (called meet). 
The maximum element and the minimum element of $\mathcal{L}$ are denoted as $\hat{\mathbf{1}}$ and $\hat{\mathbf{0}}$, respectively. 
A subset of $\mathcal{L}$ that is closed under $\vee$ and $\wedge$  forms a sublattice $\mathcal{S}$ of $\mathcal{L}$ (denoted $\mathcal{S} \leq \mathcal{L} $). 

 Let $f: \mathcal{X} \rightarrow \mathbb{R}$ be an operator function and let $g(\cdot)$ be the sum function of $f(\cdot)$ over the interval $[\hat{\mathbf{0}}, y]$. Then we have the following relationship, known as the M\"obius inversion theorem~\cite{rota1964foundations, bender1975applications}: 
\begin{equation}\label{eqn: mobius_inversion}
 g(y)=\sum \zeta(x, y) f(x)\iff f(y)=\sum_{x \leq y} \mu(x, y)g(x),
\end{equation}
where the partial order is encoded by the Zeta matrix with elements  
$\zeta(x, y) = 1$ if $x\leq y$ and 0 otherwise, $\forall x, y \in \mathcal{L}$. The inverse of $\zeta(x, y)$ is the M\"obius matrix with elements $\mu(x, y)$. 

Clearly, a measure derived by M\"obius inversion depends on both the structure of the lattice and the choice of $f$ (or $g$). Hence, M\"obius inversion on the same lattice with different operator functions can lead to different measures. Conversely, in Proposition~\ref{prop: equivalence} we show a non-trivial case where M\"obius inversion on different lattices with the same operator function leads to the same (information-theoretic) measure.

\begin{figure*}[t]
    \centering
    \includegraphics[width=0.98\textwidth]{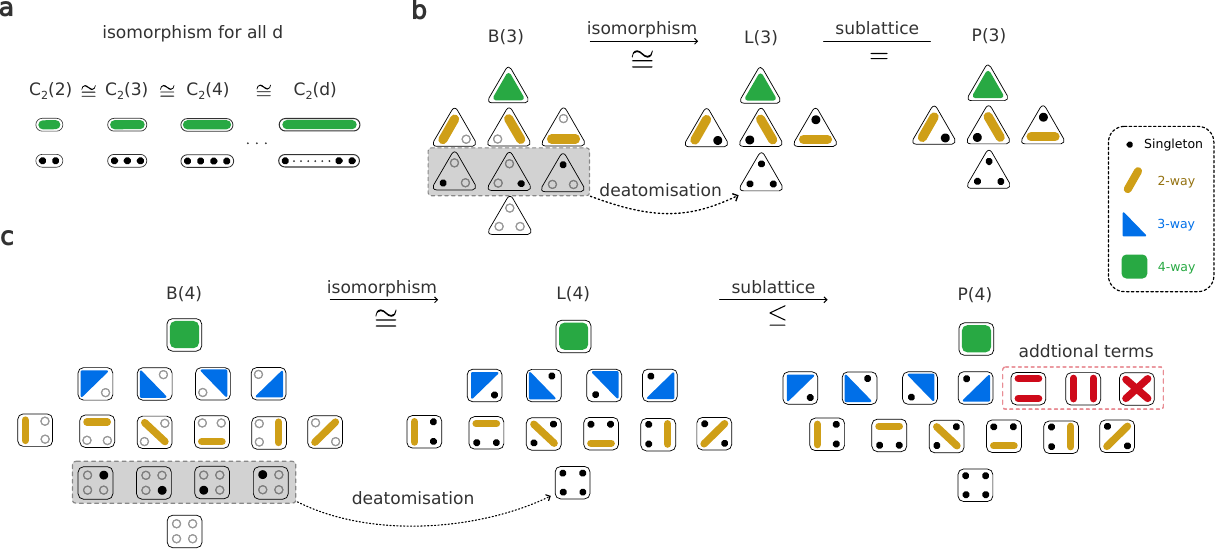}
    \caption{
    \textbf{Lattice embeddings.} The black dots indicate the marginal distributions of the singletons ($p_i$). The line, triangular and square shapes represent the joint distribution of two, three and four variables, respectively.
    (a) The two-element chain is isomorphic for all $d$. 
    (b) The deatomised $B(3)$ is isomorphic to $L(3)$, and  elements in $L(3)$ can be mapped to the non-shaded elements in $B(3)$. $L(3)$ is a sublattice of $P(3)$ (in this case equal).
    (c) For $d=4$, we see that  the deatomised $B(4)$ is isomorphic to $L(4)$, which is a (strictly smaller) sublattice of $P(4)$.
    }
    \label{fig: cartoon}
\end{figure*}

To clarify the links between different lattices it is useful to consider lattice homomorphisms~\cite{gratzer2011lattice}. A homomorphism $\varphi$ is a map of the lattice $\mathcal{L}_0$ into the lattice $\mathcal{L}_1$ which satisfies
\begin{align*}
\varphi(a \vee b)=\varphi(a) \vee \varphi(b),\qquad \varphi(a \wedge b)=\varphi(a) \wedge \varphi(b)
\end{align*}
where $a, b\in \mathcal{L}_0$ and $\varphi(a), \varphi(b)\in \mathcal{L}_1$. A one-to-one homomorphism is also called an embedding.
If $\varphi$ is a bijection, then $\varphi$ is an isomorphism (denoted $\mathcal{L}_0\cong \mathcal{L}_1$).

% Let $\mu$ be the M\"obius function of $P$~\cite{aigner1979combinatorial} p.141. For all $a, b\in P$:

Now we introduce briefly the main types of lattices that underpin the definitions of different higher-order information-theoretic measures and highlight their relationships via  lattice homomorphisms. 

First, we consider the simplest lattice, the chain, which underpins the notion of total correlation.
% \paragraph{Two-element chain.}
\begin{definition}[Chain]
    A lattice with $m$ elements is a chain, denoted $C_m$, if $a\leq b$ or $b\leq a$ for all $a,b\in C_m$, i.e. all the elements in $C_m$ are directly comparable. The length of $C_m$ is $(m - 1)$. 
\end{definition}
The simplest such lattice is the two-element chain $C_2$, consisting of elements $\hat{\mathbf{1}}$ and $\hat{\mathbf{0}}$ with M\"obius inversion
% \TL{in text?$\mu$ instead of whole?}
\begin{align}
    f(\hat{\mathbf{1}})=g(\hat{\mathbf{1}}) - g(\hat{\mathbf{0}}).
    \label{eqn: TC_mobius}
\end{align}
Let $\hat{\mathbf{1}}$ and $\hat{\mathbf{0}}$ be the joint probability density function $p_{X_1, \ldots, X_d} =: p_{1\cdots d}$ and the complete factorisation $\Pi^d_{i=1} p_{X_i}=:\Pi^d_{i=1} p_{i}$, respectively, 
and let us choose $g(\cdot) = D_{\KL}\left(\cdot \, \|\prod_{i=1}^dp_i\right) $, where $D_{\KL}(p\|q)=\int p \, \log(p/{q}) \mathrm{d}x$ is the Kullback–Leibler (KL) divergence between the probability density functions $p$ and $q$. 
Then Equation~\eqref{eqn: TC_mobius} becomes:
\begin{align}
    f(\hat{\mathbf{1}}) =& D_{\KL}\left(p_{1\cdots d}\|\prod_{i=1}^dp_i\right) - D_{\KL}\left(\prod_{i=1}^d p_i\|\prod_{i=1}^dp_i\right) \nonumber \\
    =& D_{\KL}\left(p_{1\cdots d}\|\prod_{i=1}^dp_i\right) - 0 
    = \mathrm{TC}(d),
    \label{eq:TC_lattice}
\end{align}
thus trivially recovering the total correlation $\mathrm{TC}(d)$ in Equation~\eqref{eqn: TC}.

Next, we consider the Boolean lattice $B(d)$ defined with the inclusion partial order~\cite{aigner1979combinatorial}. $B(d)$ is a natural choice for higher-order information-theoretic measures as classical information is measured in binary unit bits.
\begin{definition}[Boolean lattice]
    For a positive integer $d$, the Boolean lattice of order $d$, denoted by $B(d)$, is the poset on all the subsets of $[d] = {1, \ldots , d}$ equipped with the inclusion relation.
\end{definition}
The Boolean lattice is isomorphic to the subset lattice (the power set of a set ordered by inclusion) and the inclusion-exclusion principle appears naturally in this lattice through the M\"obius inversion
\begin{equation*}
    f(\hat{\mathbf{1}}) = \sum_{x\leq \hat{\mathbf{1}}} (-1)^{|\hat{\mathbf{1}}|-|x|} g(x).
\end{equation*}
If we choose $g(\cdot) = D_{\KL}\left(\cdot \, \|\prod_{i=1}^dp_i\right) $ as an operator function (as shown in Equation~\eqref{eq:TC_lattice})
on the Boolean lattice $B(d)$, then it follows easily that $f(\hat{\mathbf{1}})=\mathrm{II}(d)$, i.e., we recover the interaction information in Equation~\eqref{eq:II}.

The above lattices are embeddings of the partition lattice, which plays a central role in our derivations.
\begin{definition}[Partition lattice]
    Let $\mathcal{P}$ denote the set of all partitions of $\mathcal{X}$, where a partition $\pi$ is a collection of non-empty, pairwise disjoint subsets (blocks) $b_j \subseteq \mathcal{X}$ that cover $\mathcal{X}$. The partition lattice $P(d)$ is defined on $\mathcal{P}$ with the refinement ordering, i.e. for $\pi, \sigma \in \mathcal{P}$, $\pi\leq\sigma$ means every block of $\pi$ is contained in a block of $\sigma$.
\end{definition}
The M\"obius inversion on the partition lattice, given by 
\begin{equation*}
    f(\hat{\mathbf{1}}) =  \sum_{x\leq \hat{\mathbf{1}}} (-1)^{|x|-1}(|x|-1)!\, g(x)
\end{equation*}
is instrumental to define statistical measures, such as the Lancaster interaction~\cite{lancaster}, Streitberg interaction~\cite{streitberg1990lancaster, liu2023interaction}, and mixed cumulants~\cite{speed1983cumulants, mccullagh2018tensor, bonnier2024kernelized}.

The minimal ($1|2|\cdots|d$) and maximal ($12\cdots d$) elements in $P(d)$ form a sublattice denoted as $C_2(d)$, which is isomorphic to the two-element chain $C_2$ for any $d$ (Figure~$\ref{fig: cartoon}$a). 
Hence the $C_2$-induced measure $\mathrm{TC}(d)$  always measures the difference between the top of the lattice, $\hat{1}$, and the bottom of the lattice, $\hat{0}$, i.e., between the joint distribution and the product of all the marginals, independently of $d$. As $d$ grows, the lattice $C_2(d)$ remains static (just two elements) and $\mathrm{TC}$ becomes progressively less informative.

Another sublattice of the partition lattice that is of interest here is the sublattice where partitions contain at most one non-singleton block. We denote this sublattice as the Lancaster lattice $L(d)$ (see Figure~$\ref{fig: cartoon}$b,c). 
\begin{lemma}\label{lemma: 1}
    % The Lancaster lattice $L(d)$ is an embedding of the Boolean lattice $B(d)$. More specifically, 
    $L(d)$ is isomorphic to $B(d)$ with the exclusion of singleton elements (deatomisation).
\end{lemma}
\begin{proof} 
For a proof, see Appendix~\ref{pf: lemma1}. 
\end{proof}
As a consequence, $B(d)$-based measures can also be generated from $L(d)$, a sublattice of $P(d)$, e.g., the computation of the Lancaster interaction~\cite{lancaster,liu2023interaction}.

\textit{\underline{Remark} (Simplicial complexes):}
A popular representation of higher-order systems is through simplicial complexes~\cite{bianconi2021higher}. 
The elements in a ($d-1$)-simplex with inclusion ordering form a lattice isomorphic to the Boolean lattice $B(d)$.
Furthermore, the boundary matrices of a ($d-1$)-simplex appear as block submatrices in the Zeta matrix and the M\"obius matrix of the partition lattice $P(d)$ (see Appendix~\ref{sec: simplicial_complex}).

%\section{Proposed measure (new)}
\section{Streitberg Information}\label{sec: 4}

Following our discussion above, it is then natural to define 
the following information-theoretic measure based on the M\"obius inversion of the full partition lattice $P(d)$ with an operator function given by some divergence function acting on probability distributions.

\begin{definition}\label{def: Li_SI}
    The Streitberg information $\mathrm{SI}(d)$ is then obtained from the M\"obius inversion of P(d) with operator function $g( \cdot) = D(\cdot\|\prod_{i=1}^{d}p_i)$:
    \begin{align}
        \mathrm{SI}(d) =  \sum_{\pi\in P(d)} (|\pi|-1)!(-1)^{|\pi|-1} D\left(p_{\pi}\|\prod_{i=1}^{d}p_i\right) \label{eqn: SI}
    \end{align}
    where 
    $p_{\pi} = \prod_{j=1}^r p_{b_j}$ is the corresponding factorisation with respect to $\pi = b_1\vert b_2\vert\ldots\vert b_r$, and  $D(\cdot \| \cdot)$ is a well-defined divergence between probability functions. 
\end{definition}

Similarly, one can define the Lancaster information restricted to the Lancaster (sub)lattice $L(d)$:
\begin{equation}
        \mathrm{LI}(d) = \sum_{\pi_l\in P(d)} (-1)^{|\pi_l|-1} D\left(p_{\pi_l}\|\prod_{i=1}^{d}p_i\right)
        \label{eq:LI_def}
\end{equation}
where $\pi_l$ are the partitions with at most one non-singleton block.

A direct consequence of our framework is that the (typical) choice of the KL divergence as the operator function leads to the equivalence of measures derived from different lattices. This is due to cancellations of interactions that lead to a collapse of the partition lattice onto particular sublattices leading to loss of information by disregarding particular interactions. In particular, it is easy to prove that, if we choose the KL divergence as the operator function $g( \cdot) = D_\text{KL}(\cdot\|\prod_{i=1}^{d}p_i)$ as in Eq.~\eqref{eq:TC_lattice}, then interaction information, Lancaster information and Streitberg information are all equivalent, despite being derived from M\"obius inversions on distinct lattices.
\begin{proposition}\label{prop: equivalence}
When $D(\cdot\|\cdot)$ in Equation~\eqref{eqn: SI} is the KL divergence, then we have the following equivalences:
\begin{equation}
    \mathrm{II}(d) = \mathrm{LI}(d) = \mathrm{SI}(d) \qquad \text{for } d\geq 2.
    \label{eq: equality}
\end{equation}
\end{proposition}
\begin{proof} 
The first equality can be proved by counting the number of double-counted marginals due to the KL formalisation. For the second equality, it suffices to show that the KL divergence associated with the non-singleton partitions (except $\hat{\mathbf{1}}$) vanish. We prove this in three ways in Appendix~\ref{SI: prop} using (i) the information geometry of KL divergence, (ii) interval lattice, and (iii) the relationship between Lancaster information and Streitberg information.
\end{proof}

The equivalence of M\"obius functions for different lattices under KL divergence shown in  Equation~\eqref{eq: equality} is the result of certain interaction terms cancelling out when KL is chosen as the operator function. This motivates our considering alternative divergences with different information geometry.
Here, we focus on the Tsallis-Alpha divergence $D_\alpha(p \| q)$~\cite{tsallis1988possible}, a classic generalisation of the KL divergence
\begin{equation}\label{eq: D_ta}
    D_\alpha(p \| q)=\frac{1}{\alpha-1} \left(\int_x p^{\alpha}(x) q^{1-\alpha}(x) \mathrm{d} x \, - \, 1\right),
\end{equation}
which has found a wide range of applications in machine learning~\cite{wang2021variational,amid2019two,vila2011tsallis}.
This divergence is uniquely characterised by the parameter $\alpha$: it reduces to the KL divergence when $\alpha=1$, to the Hellinger distance when $\alpha=0.5$~\cite{hellinger1907orthogonalinvarianten}, and to the $\chi^2$ distance when $\alpha=2$~\cite{gibbs2002choosing}. The Tsallis-Alpha divergence is equivalent to the Amari-Alpha divergence up to a substitution of $\alpha$~\cite{amari2001information}. Crucially, $D_\alpha(p \| q)$ only decomposes into lower-order terms when $\alpha=1$, i.e., when it becomes the KL divergence~\cite{morales2021generalization, huang2016generalization}.

As a result, when $D(\cdot\|\cdot)$ in Equation~\eqref{eqn: SI} is the Tsallis-Alpha divergence, $\mathrm{SI}(d)$ vanishes if the joint distribution can be factorised in any way. In contrast, when $D(\cdot\|\cdot)$ in Equation~\eqref{eqn: SI} is the KL divergence, we have equivalence to $\mathrm{LI}(d)$, which  vanishes if the factorisation contains at least one singleton block (Figure~\ref{fig: synthetic}a-b). Henceforth, we focus on $\mathrm{SI}(d)$ with Tsallis-Alpha divergence, to take advantage of the factorisations in the full  partition lattice.

\underline{\textit{{Remark:}}} When choosing $\alpha$ outside the range of $(0, 1)$, it has been shown that the Tsallis-Alpha divergence can become unbounded~\cite{poczos2011estimation}, and similarly for the Renyi divergence, which is closely related to the Tsallis-Alpha divergence~\cite{van2014renyi}. Hence we restrict our numerical experiments to the range $\alpha \in (0, 1)$.

\paragraph{Recursiveness}
Measures on lattices are generated by the M\"obius function through inversion of the Zeta function. Here we show that, based on Equation~\eqref{eqn: mobius_inversion}, the expressions of $\mathrm{SI}(d)$ in Equation~\eqref{eqn: SI} can be expressed in terms of sums of lower-order $\mathrm{SI}(d)$, and similarly for $\mathrm{LI}(d)$ in Equation~\eqref{eq:LI_def}:
\begin{lemma}\label{lemma: recurisve}
    \begin{align}\label{eqn: Zeta2}
   \mathrm{SI}(d) &= D_{\alpha}\left(p_{1\ldots d}\|\prod_{i=1}^d p_i\right)-\sum_{\pi\setminus \hat{\mathbf{1}}} \mathrm{SI}(\pi),\\
   \mathrm{LI}(d) &=  D_{\alpha}\left(p_{1\ldots d}\|\prod_{i=1}^d p_i\right)-\sum_{\pi_l\setminus \hat{\mathbf{1}}} \mathrm{LI}(\pi_l). 
\end{align}
\end{lemma}
\begin{proof}
This follows from rearranging the Zeta functions (see Appendix~\ref{pf: lemma2}).
\end{proof}
\underline{\textit{{Remark:}}} A generalised higher-order information defined on the partition $\pi$, $\mathrm{SI}(\pi)$, can be understood using the interval lattice
$[\hat{\mathbf{0}}, \pi]$, e.g., $\mathrm{SI}(12|34) = D_{\alpha}(p_{12}p_{34}\|p_1p_2p_3p_4) - D_{\alpha}(p_{12}\|p_1p_2) - D_{\alpha}(p_{34}\|p_3p_4)$. 
This is always zero when $\alpha=1$ due to KL decomposition, hence the equivalence in Proposition~\ref{prop: equivalence}. See Appendix~\ref{SI: generalised_interaction} for more details. 

In this sense, the Zeta functions of $\mathrm{SI}(d)$ and $\mathrm{LI}(d)$ can be seen as attempts to recursively decompose the divergence between the joint distribution and the product of marginals. This formulation gives an intuitive interpretation of the vanishing condition $\mathrm{SI}(d)=0$: the dependence in the $d$-order system can be completely explained by the lower order terms. 
It also follows from this rewriting that the sign of
$\mathrm{SI}(d)$ and $\mathrm{LI}(d)$ depends on the difference between the $D_{\KL}\left(p_{1\cdots d}\|\prod_{i=1}^dp_i\right)$ and the sum of the respective lower order information terms.

\underline{\textit{{Remark:}}} The recursiveness property underpins the definition of connected information~\cite{schneidman2003network}, where it was used without explicit links to lattice theory. The terms in the connected information are elements of the power set of $\mathcal{X}$; hence it is directly comparable to information-theoretic measures derived from the Lancaster and Boolean lattices, instead of  
$\mathrm{SI}(d)$, which is defined on the complete partition lattice.

\paragraph{Symmetry}
Both $\mathrm{SI}(d)$ and $\mathrm{LI}(d)$ are invariant under permutation of the variables. This contrasts with some information-theoretic measures such as PID~\cite{williams2010nonnegative} and PED~\cite{ince2017partial} that are directed, i.e., they partition the variables into a source set and a target set. In complex multivariate systems, computing a directed measure requires either looping through all the subsets of variables or constructing a justifiable target set, both of which can be challenging or even unfeasible. Moreover, the decomposed information atoms obtained through PID are target-specific and thus insufficient for a comprehensive description of the relationships between variables.

\paragraph{Emergence}
A non-zero value of $\mathrm{SI}(d)$ indicates the presence of a $d$-order interaction. A special case of this condition is when there are no lower-order interactions, i.e., if $\mathrm{SI}(d^*)=0$ for all $2\leq d^* \leq d-1$. We denote these as \textit{emergent higher-order interactions}~\cite{liu2023kernel}, since the higher-order interaction is  explained solely by the dependence among $d$ variables given that the only remaining term in Equation~\eqref{eqn: Zeta2} is $D(p_{1\ldots d}\|\prod_{i=1}^d p_i)$.

\paragraph{Consistent Estimator}
In many real-world scenarios, the distribution function of the data is unknown, making it impossible to compute information-theoretic measures directly. In this scenario, non-parametric estimators that can reliably approximate the true values of our measures are needed.
Here, we avoid direct density estimation and consider asymptotically unbiased estimators based on the $L_2$-consistent estimation of Tsallis-Alpha divergence with $k$-nearest-neighbour ($k$NN) statistics~\cite{poczos2011estimation} (implemented using python package ITE~\cite{szabo2014information}).
The resulting estimators are a linear sum of the divergences and hence consistent by Slutsky's theorem~\cite{chung2001course}.

Notice in Equation~\eqref{eqn: SI} that each divergence term consists of a factorisation and its respective complete factorisation, but the singletons in any factorisation cancel with the singletons in the complete factorisation, e.g., $D_{\alpha}(p_1p_2p_{34}\|p_1p_2p_3p_4) = D_{\alpha}(p_{34}\|p_3p_4)$. This makes this latter example a 2- rather than a 4-dimensional $k$NN problem and leads to more efficient computations (see Appendix~\ref{SI: compute} for more computational details). As a result, summing over all partitions reduces to a sum over the power set and the partitions with no singletons. 
To approximate the distributions of the complete factorisation, we permute the realisations of each variable randomly to break any dependence among the variables, e.g., the realisations from $p_1p_2p_3p_4$ are generated by randomly permuting the realisation from $p_{1234}$.

\begin{figure*}[htb!]
    \centering
    \includegraphics[width=0.95\textwidth]{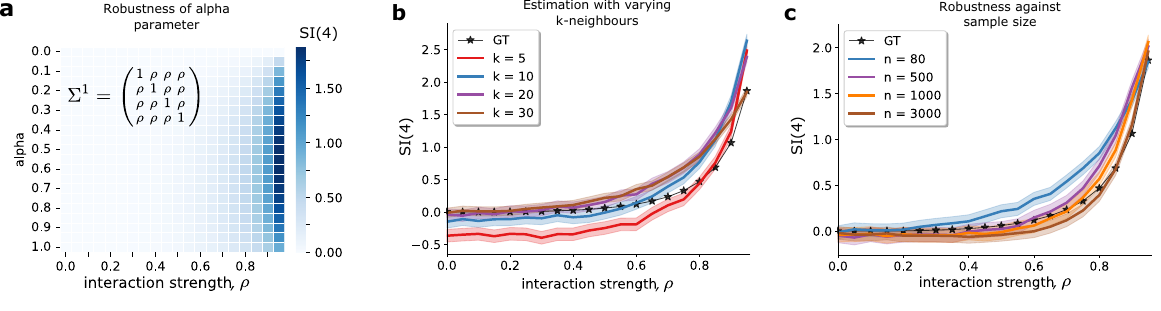}
    \caption{\textbf{Robustness of Streitberg information ($d=4$).}
    All calculations in this figure are for data from a MVG with covariance matrix $\Sigma^1 (\rho)$, where $\rho$ indicates the interaction strength.
     (a) The behaviour of $\mathrm{SI}(4)$ as a function of $\rho$ is consistent across $\alpha$. 
    (b) Increasing the number of neighbours $k$ improves the estimation accuracy.
    (c) Increasing the sample size $n$ improves the estimation accuracy for a fixed number of neighbours ($k=5$). 
   }
    \label{fig: estimation}
\end{figure*}

\section{Synthetic Experiments}\label{sec: Synthetic experiments}
To validate the Streitberg information, we use synthetic data from multivariate Gaussians (MVG) with different ground truth interactions, as defined by their correlation structure, as well as XOR and COPY gates. Since the advantages of  Streitberg information become apparent for $d>3$ (see Figure~\ref{fig: cartoon}c),
we show here results for $d=4$.
Similar, consistent conclusions for experiments with $d=3$ and $d=5$ are presented in Appendix~\ref{SI: order35_experiments}. For the analytical solution of MVG, see Appendix~\ref{sec: tsallis}.

\paragraph{Varying the Alpha Parameter of the Divergence}
We begin by exploring the behaviour of $\mathrm{SI}(4)$ as we vary the parameter $\alpha$ of the Tsallis-Alpha Divergence $D_\alpha( \cdot \| \cdot)$.  We use data from a MVG with variance $\Sigma^1$ where $0 \leq \rho < 1$ defines the interaction strength between variables (see inset of Fig.~\ref{fig: estimation}a).
As discussed above, $\alpha \in (0,1)$ tunes the properties of the divergence with particular cases at $\alpha=0.5$ (Hellinger distance) and $\alpha=1$ (KL divergence). Our numerics in Fig.~\ref{fig: estimation}a show that varying $\alpha$ has little effect on the relationship with $\rho$, highlighting the robustness of $\mathrm{SI}(d)$. The $\mathrm{SI}(4)$ is most sensitive for $\alpha=0.5$, probably because $D_\alpha( \cdot \| \cdot)$ is symmetric in its argument only when $\alpha=0.5$. We thus fix $\alpha=0.5$ for all analyses below.
\begin{figure*}[t]
    \centering
    \includegraphics[width=.97\textwidth]{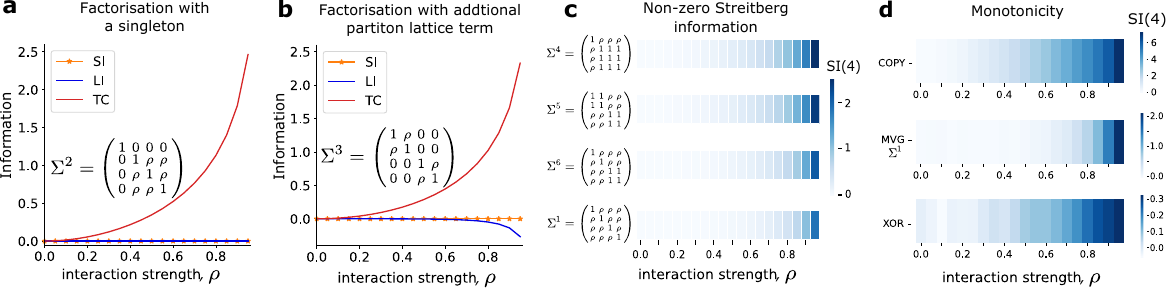}
    \caption{\textbf{Validation of Streitberg information.}
    (a) $\mathrm{TC}(4)$ fails to vanish for $p_{1}p_{234}$, while
    (b) both $\mathrm{TC}(4)$ and $\mathrm{LI}(4)$ fail to characterise $p_{12}p_{34}$. $\mathrm{SI}(4)$ correctly vanishes in both cases.
    (c) The magnitude of $\mathrm{SI}(4)$ is influenced by the extent to which the joint distribution can be factorised. 
    (d) Streitberg Information exhibits monotonic behaviour consistently across varying types of interaction, and the magnitude indicates again the difficulty of factorising the joint distribution. 
    }
    \label{fig: synthetic}
\end{figure*}

\begin{figure*}[t]
    \centering
    \includegraphics[width=0.98\textwidth]{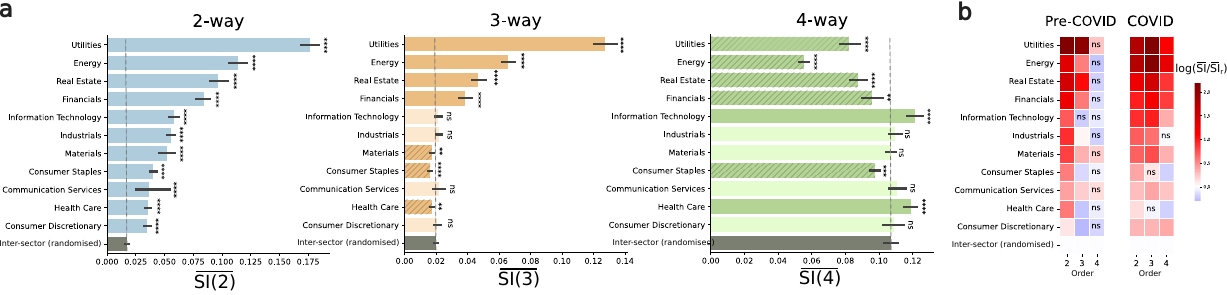}
    \caption{
    \textbf{$d$-order Streitberg information in stock returns vary within and across sectors.}
    (a) $\mathrm{SI}(d)$ was computed between daily returns of stocks from 2010-2024 within sectors (coloured bars) and across random sectors as baseline (grey bars) for $d=2,3,4$. Each bar represents the average magnitude of $\mathrm{SI}(d)$ across $500$ samples of $d$ stocks. Bars with deeper colours (resp. shaded) have significantly larger (resp. smaller) $\mathrm{SI}(d)$ than the inter-sector grey bars. Bars with pale colours display non-significant differences with respect to the inter-sector baseline.  
    (b) Log ratio between the $\mathrm{SI}(d)$ within each sector and the random baseline across sectors for stock returns pre- and post-January 2020 (COVID onset). We report significance of two-sample t-test of $\mathrm{SI}(d)$ within sector vs. between sector  (*, $p<0.05$; **, $p<0.01$; ***, $p<0.001$; ****, $p<0.0001$; n.s., not significant). There is an increase in significantly different values of $\mathrm{SI}(4)$ post-COVID across sectors. 
    }
    \label{fig: sp500}
\end{figure*}

\paragraph{Estimation Accuracy}
To assess the robustness of the estimation of Streitberg information, we varied (i) the number of neighbours $k$ for $k$NN-based density estimation and (ii) the number of samples $n$.
First, we generated $n=80$ $iid$ realisations of a MVG with covariance $\Sigma^1$ and varied the value of $k$. Even with as few as $k=5$ neighbours, the estimation of the ground truth is reasonably accurate, albeit with larger estimation errors for $\rho=0$ (Figure~\ref{fig: estimation}b).
Next, we fixed $k=30$ and varied the number of samples $n$. The estimations are relatively accurate even with only $n=80$ with improved estimation for all $\rho$ as $n$ is increased (Figure~\ref{fig: estimation}c).

\paragraph{Vanishing Condition}
Next, we computed $\mathrm{SI}(4)$, $\mathrm{LI}(4)$ and $\mathrm{TC}(4)$ for varying factorisations of $p_{1234}$: (i) $\Sigma^2$ where $p_{1234}=p_1p_{234}$, and (ii) $\Sigma^3$ where $p_{1234}=p_{12}p_{34}$. A well-behaved higher-order information-theoretic measure should stay at zero when a lower-order factorisation of the joint distribution is present.
As expected, $\mathrm{TC}(4)$ fails to capture either factorisation, whilst $\mathrm{LI}(4)$ is only able to detect the factorisation consisting of singletons. Importantly, $\mathrm{SI}(4)$ is able to detect both factorisations (Figure~\ref{fig: synthetic}a-b).

\paragraph{Interpretation of $\mathrm{SI}(d)$  }
Now, we illustrate the interpretation of $\mathrm{SI}(d)$ when it is non-zero. We define four different MVGs using the covariance matrices $\Sigma^4$, $\Sigma^5$, $\Sigma^6$ and $\Sigma^1$ in Figure~\ref{fig: synthetic}c
which are equivalent when $\rho$ tends to 1 and represent block matrices with factorisable distributions when $\rho=0$.
Figure~\ref{fig: synthetic}c shows that $\mathrm{SI}(4)$ increases more rapidly as a function of the interaction strength $\rho$ when the covariance matrix contains more independent sub-blocks (in the factorisable limit of $\rho=0$). For covariance matrices $\Sigma^4$ and $\Sigma^5$, which have the same number of factorisable sub-blocks in the limit, $\mathrm{SI}(4)$ increases more slowly for $\Sigma^5$. 
This difference can be explained by the structure of the probability distributions: pulling out one singleton from $p_{12}p_{34}$ creates two new singletons, whereas doing the same from $p_{1}p_{234}$ results in at most one new singleton.
Therefore the MVG with $\Sigma^5$ is more amenable to further factorisation. 
This suggests that a larger $\mathrm{SI}(d)$ captures a lower likelihood of factorisation, pointing to stronger dependency ties within the set of variables.

\paragraph{Monotonicity} % n = 1000
Finally, we investigate the monotonicity of $\mathrm{SI}(d)$ using data sets from an XOR gate, a COPY gate and MVG. The MVG is generated with covariance matrix $\Sigma^1$. For the XOR and COPY gates, we proceed as follows. We generate $n$ $iid$ samples of $W, X, Y, Z {\sim} \mathcal{U}(0,4)$. For the XOR gate, we set $Z_{:i} = (W_{:i}+X_{:i}+Y_{:i})\mod 4$ and  $Z_{i+1:n} \sim \mathcal{U}(0,4))$. For the COPY gate, we set $W_{:i} = Z_{:i}$, $X_{:i} = Z_{:i}$, $Y_{:i} = Z_{:i}$. We then gradually increase the interaction proportion, $0 \leq i/n \leq 1$. 
We find that $\mathrm{SI}(d)$ increases monotonically with the interaction strength in all three examples (Figure~\ref{fig: synthetic}d).
By construction, the XOR data set does not contain pairwise or 3-way interactions, 
and breaking the 4-way interaction makes all the variables into singletons. 
In contrast, the COPY and MVG data sets contain pairwise, 3- and 4-way interactions when $i$ and $\rho$ are non-zero. In comparison to MVG, the COPY data set contains stronger interactions since the variables are eventually identical ($\rho=1$), and thus the magnitude of $\mathrm{SI}(d)$ is greater than MVG.

\section{Applications}
Finally, we exemplify $\mathrm{SI}(d)$ on applications in finance, neuroscience and machine learning.

\paragraph{Interactions in Stock Market Returns}
While traditional portfolio optimisation frameworks, which rely on limiting the covariance of investments~\cite{vermorken2011gics}, provide valuable insights into risk and return, they often overlook higher-order interactions among stocks. 
Here, we compute the Streitberg information between stocks in the S\&P 500 using daily returns (assumed as $iid$~\cite{ali1982identical}) from 4 Jan 2010 to 24 Apr 2024. For comparison, we computed both intra- and inter-sector pairwise, 3-, and 4-way Streitberg information (see Appendix~\ref{sec: stockmarket} for details).

As expected, results in Figure~\ref{fig: sp500}a show that intra-sector pairwise interactions are significantly stronger than inter-sector (grey bar) across all sectors owing to, e.g., shared economic drivers or supply chain interdependencies. At 3-way, however, we observe variation across sectors; sectors like Utilities which are heavily regulated
exhibited increased 3-way interactions, whilst sectors such as Health Care which are more diverse (including pharma companies, biotech firms and healthcare providers) displayed less 3-way interactions. 
We see further differences at 4-way Streitberg information. For example, the energy sector which is primarily driven 
by commodity prices, geopolitical events and regulations, had lower 4-way interactions. 
On the other hand, Information Technology, a sector that is characterised by rapid innovation and interdependent product ecosystems, displayed higher 4-way interactions.
We further computed the Streitberg information before and after 2 Jan 2020 to study the impact of COVID on the interdependencies between stocks (Figure~\ref{fig: sp500}b). Pre-COVID, we found no significant difference in intra-sector 4-way interactions relative to inter-sector; however, since COVID almost all sectors now show significantly different 4-way interactions (mostly higher, some lower) relative to inter-sector.

\begin{figure}[htb!]
\centering
\label{fig: monkey}
\includegraphics[width=0.45\textwidth]{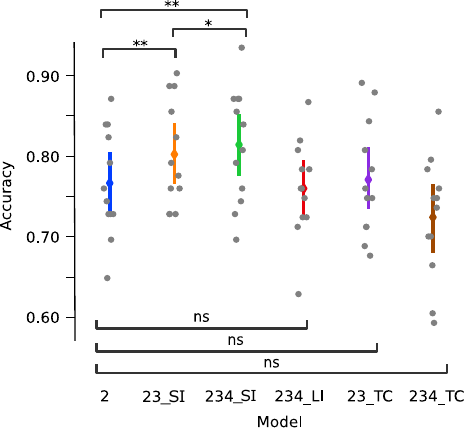}
\caption{\textbf{Streitberg information significantly improves the decoding of preparatory stage \textit{vs.} motor action in neural activity.} Classifiers that include  higher-order features from $\mathrm{SI}$ , $\mathrm{LI}$ and $\mathrm{TC}$ are trained to identify trial stage (preparation \textit{vs.} action). We report the significance of paired t-tests between model accuracies across sessions (*, $p<0.05$; **, $p<0.005$; ns, not significant). Adding SI features ($d=3,4$) improves classifier accuracy, whereas adding higher-order LI and TC features induces no significant improvement in accuracy compared to the pairwise model (`2'). }
\end{figure}
\paragraph{Decoding Neural Spiking Data}
Prior work has found higher-order interactions to play an important role in neuroscientific data~\cite{schneidman2003synergy, luppi2022synergistic, luff2024neuron}. Therefore, we asked whether higher-order interactions improved decoding of electrophysiological recordings of a monkey performing a delayed centre-out hand-reaching task~\cite{gosztolai2025marble, pandarinath2018inferring}. In each recording session, the monkey was instructed to perform seven fixed hand movements for a randomly defined number of trials. 
For a given session, we exhaustively computed the 2-, 3- and 4-way Streitberg information, Lancaster information and total correlation of all possible combinations across a set of 12 neurons. This was repeated for each time point of every condition across 12 sessions (details in Appendix~\ref{SI:monkey}).

We asked whether higher-order interactions were more important during the preparatory stages or during motor action. We trained a logistic classifier using varying orders of $\mathrm{SI}$, $\mathrm{LI}$ and $\mathrm{TC}$ as features. To avoid overfitting due to the increasing number of features as the order increases, we applied PCA to the $d$-order interaction features and used a constant number of principal components for all orders.
Our analysis revealed that while incorporating 3-way $\mathrm{TC}$ does not necessarily enhance accuracy, the inclusion of 3-way $\mathrm{SI}$, which is equal to $\mathrm{LI}$ for $d=3$, significantly boosts model accuracy. When we increased the order to 4, the Streitberg information once again improves the classification accuracy of trial stages, while the additional $\mathrm{TC}$ and $\mathrm{LI}$ fail, highlighting the importance of $\mathrm{SI}$ in neural coding.

\paragraph{Feature Selection}

Feature selection is a critical step in machine learning, aiming to identify the most relevant variables that contribute to accurate predictions and model interpretability. Traditionally, feature selection techniques focus on assessing individual features based on their predictive power or statistical dependence with the target variable $Y$~\cite{poczos2012copula}.
However, such methods inherently overlook higher-order interactions between features---interactions that can be crucial in capturing the underlying structure of complex data sets---and can therefore lead to incorrect interpretations and feature relevance.

\begin{table}[htb!]
\centering
\caption{\textbf{Comparison of Streitberg information and benchmark feature selection methods.} Only Streitberg information identifies the ground truth variables ($\CIRCLE$) that generate $Y$. Other methods either include confounding variables ($\textcolor{red}{\CIRCLE}$) or miss variables in the generating set ($\textcolor{red}{\Circle}$) and lead to larger errors (MSE).}
\vspace{5pt}
\label{tab: feature_selection}
\resizebox{0.45\textwidth}{!}{%
\begin{tabular}{lcccccc}
\hline
Methods & $X_1$ & $X_2$ &  $X_3$ & \textcolor{red}{$X_4$} & \textcolor{red}{$X_5$} & MSE \\ \hline
Variance           & $\CIRCLE$ & $\CIRCLE$ & $\CIRCLE$ & \textcolor{red}{$\CIRCLE$} & \textcolor{red}{$\CIRCLE$} & 0.00054  \\
Univariate         & $\CIRCLE$ & $\CIRCLE$ & $\CIRCLE$ & \textcolor{red}{$\CIRCLE$} & \textcolor{red}{$\CIRCLE$} & 0.00054 \\
Sequential         & $\CIRCLE$ & $\CIRCLE$ & $\textcolor{red}{\Circle}$   &              &              & 0.09921 \\
RFE                & $\CIRCLE$ & $\CIRCLE$ & $\textcolor{red}{\Circle}$   &              &              & 0.09921  \\
Pairwise MI        & $\CIRCLE$ & $\textcolor{red}{\Circle}$   & $\textcolor{red}{\Circle}$   &              &              & 0.12521  \\
SHAP               & $\CIRCLE$ & $\CIRCLE$ & $\CIRCLE$ & \textcolor{red}{$\CIRCLE$} & \textcolor{red}{$\CIRCLE$} & 0.00054  \\
$\mathrm{TC}$                 & $\CIRCLE$ & $\CIRCLE$ & $\CIRCLE$ & \textcolor{red}{$\CIRCLE$} & \textcolor{red}{$\CIRCLE$} & 0.00054  \\
$\mathrm{LI}$                 & $\CIRCLE$ & $\CIRCLE$ & $\CIRCLE$ & \textcolor{red}{$\CIRCLE$} & \textcolor{red}{$\CIRCLE$} & 0.00054  \\
\textbf{$\mathrm{SI}$ (ours)} & $\CIRCLE$ & $\CIRCLE$ & $\CIRCLE$ &              &              & \textbf{0.00002}  \\
  \hdashline
\textit{Ground truth}                 & $\CIRCLE$ & $\CIRCLE$ & $\CIRCLE$ &              &              & \textit{0.00002}  \\ \hline
\end{tabular}%
}
\end{table}
Here, we construct a dataset with five independent variables  $X_1, X_2, X_3, X_4, X_5$ and define a target variable $Y$ as the XOR of $X_1, X_2, X_3$. Here $X_4, X_5$ are MVG with covariance $[[1, 0.95], [0.95, 1]]$. The ground truth (GT) interaction between these five variables is $\{Y, X_1, X_2, X_3\}\independent\{X_4, X_5\}$, such that no single variable or pair of variables is predictive of $Y$. 

In Table~\ref{tab: feature_selection}, we compare the performance of commonly used algorithms, such as recurrent feature selection~\cite{sklearn} and SHAP~\cite{SHAP}, with methods based on higher-order information-theoretic measures. The set of GT features that explain $Y$ are exactly found with Streitberg Information, yielding the least mean squared error (MSE). Alternative methods select the wrong features since they are unable to identify the complex higher-order interactions between the input features and the target variable. $\mathrm{TC}$ and $\mathrm{LI}$ also fail as they cannot detect the partial factorisation in our GT. The methods that select less than three variables perform poorly, 
while the methods that select all five variables such as SHAP have comparably low MSE, but are both computationally more expensive and lack explainability since they select all features when only three should be selected.

\section{Discussion}\label{sec: discussion}
Our work offers a rigorous and systematic approach for the  detection of higher-order interactions using information-theoretic measures, but has some limitations. 
While the sign of $\mathrm{SI}(d)$ is determined by the difference between the total dependence and the sum of the lower-order terms through the recursiveness property, how it changes with order $d$ remains unexplored. 
Moreover, although the magnitude of the Streitberg information can be related to the factorisability of the joint probability distribution, the bounds of the measure across $\alpha$ remain unclear.

A key motivation for asymmetric measures is their potential to provide insights into causality, but the link between our Streitberg information measure and causality would need further exploration. Despite the fundamental differences between our measure and PID, 
Streitberg information is more scalable compared with PID. This can be understood through our lattice-theoretical framework, which also provides theoretical insights into the formulation of PID via the free distributive lattice and its scaling with the number of variables (see Appendix~\ref{SI: PID}).  This will be the focus of future work.

Other limitations include the fact that the non-parametric estimation of the Streitberg information assumes that the data is $iid$, a strong assumption in many real-world data sets, especially those that contain time series. Also the $k$NN-based estimation typically does not scale well with sample size or data dimensionality~\cite{chen2024mutual}, and although recent work has explored using neural networks to estimate mutual information~\cite{belghazi2018mutual, xu2024neural}, as of now, there are no similar estimators for Tsallis-Alpha divergence. 

There are several open directions for future work.
Specifically, the investigation of the sign and bounds of Streitberg information across order $d$ and $\alpha$; how $\mathrm{SI}$  can be accurately approximated when the data has temporal dependence; the connection between Streitberg information and causal discovery; linking the lattice-based framework with the formalisation of directed measures that split the variables into source and target sets; and defining more scalable and flexible estimators for Streitberg information.
\section*{Acknowledgements}
MB acknowledges support by EPSRC through grant EP/W024020/1 funding the project ``Statistical physics of cognition'' and grant EP/N014529/1 funding the EPSRC Centre for Mathematics of Precision Healthcare at Imperial, and by the Nuffield Foundation under the project ``The Future of Work and Well-being: The Pissarides Review''.
RP acknowledges funding from the Deutsche Forschungsgemeinschaft (DFG, German Research Foundation) Project-ID 424778381-TRR 295.

The authors would like to thank Pedro Mediano, Jianxiong Sun and Xing Liu for valuable discussions.

\newpage
\bibliographystyle{unsrt}
\bibliography{ref}

 \begin{enumerate}

\clearpage
 \item For all models and algorithms presented, check if you include:
 \begin{enumerate}
   \item A clear description of the mathematical setting, assumptions, algorithm, and/or model. [Yes]
   \item An analysis of the properties and complexity (time, space, sample size) of any algorithm. [Yes]
   \item (Optional) Anonymized source code, with specification of all dependencies, including external libraries. [Yes]
 \end{enumerate}

 \item For any theoretical claim, check if you include:
 \begin{enumerate}
   \item Statements of the full set of assumptions of all theoretical results. [Yes]
   \item Complete proofs of all theoretical results. [Yes]
   \item Clear explanations of any assumptions. [Yes]     
 \end{enumerate}

 \item For all figures and tables that present empirical results, check if you include:
 \begin{enumerate}
   \item The code, data, and instructions needed to reproduce the main experimental results (either in the supplemental material or as a URL). [Yes]
   \item All the training details (e.g., data splits, hyperparameters, how they were chosen). [Yes]
         \item A clear definition of the specific measure or statistics and error bars (e.g., with respect to the random seed after running experiments multiple times). [Yes]
         \item A description of the computing infrastructure used. (e.g., type of GPUs, internal cluster, or cloud provider). [Yes]
 \end{enumerate}

 \item If you are using existing assets (e.g., code, data, models) or curating/releasing new assets, check if you include:
 \begin{enumerate}
   \item Citations of the creator If your work uses existing assets. [Yes]
   \item The license information of the assets, if applicable. [Not Applicable]
   \item New assets either in the supplemental material or as a URL, if applicable. [Not Applicable]
   \item Information about consent from data providers/curators. [Not Applicable]
   \item Discussion of sensible content if applicable, e.g., personally identifiable information or offensive content. [Not Applicable]
 \end{enumerate}

 \item If you used crowdsourcing or conducted research with human subjects, check if you include:
 \begin{enumerate}
   \item The full text of instructions given to participants and screenshots. [Not Applicable]
   \item Descriptions of potential participant risks, with links to Institutional Review Board (IRB) approvals if applicable. [Not Applicable]
   \item The estimated hourly wage paid to participants and the total amount spent on participant compensation. [Not Applicable]
 \end{enumerate}

 \end{enumerate}

\appendix
\onecolumn

\section{Proofs}

\subsection{Proof of Lemma~\ref{lemma: 1}}\label{pf: lemma1}
\begin{proof}
    One can observe that the structure of the $B(d)$ is identical to $L(d)$ up to isomorphism except on the second last level on which the singleton elements lie. Two examples for $d=3,4$ are shown in Figure~\ref{fig: cartoon}b-c. The length of $B(d)$ minus the length of $L(d)$ is precisely one and this correspond to the level of singletons which are united together to form $\hat{\mathbb{0}}$ in $L(d)$. Excluding these elements leads to the isomorphism.
\end{proof}

\subsection{Proof of Proposition~\ref{prop: equivalence}}\label{SI: prop}
\begin{proof}
We first show that the first equality ($\mathrm{LI}(d) = \mathrm{II}(d)$) in Equation~\ref{eq: equality} holds.
In $\mathrm{LI}(d)$, we sum the divergences over $\pi_l$, the partitions with at most one non-singleton block. In each divergence term, the singletons in the probability distribution function corresponding to $\pi_l$ can be cancelled with the singletons present in the complete factorisation $\prod_{i=1}^d\mathbb{P}_i$. The resulting divergence is measuring the difference between the non-singleton block and its corresponding complete factorisations, i.e., given $\mathbb{P}_{\pi_l} = \mathbb{P}_{NS}\prod_{j\in S}\mathbb{P}_{j}$ where $\mathbb{P}_{NS}$ represents the joint distribution of the non-singleton block and $\mathbb{P}_{j}$ are the distributions of the singletons, 
\begin{equation*}
    D_{\KL}\left(\mathbb{P}_{\pi_l}\|\prod_{i=1}^d\mathbb{P}_i \right)=
    D_{\KL}\left(\mathbb{P}_{NS}\prod_{j\in S}\mathbb{P}_{j}\|\prod_{j\in S}\mathbb{P}_{j}\prod_{k\in NS}\mathbb{P}_{k} \right)= D_{\KL}\left(\mathbb{P}_{NS}\|\prod_{k\in NS}\mathbb{P}_{k} \right)
\end{equation*}
The set $NS$ forms the powerset of $d$ elements, hence the terms in $\mathrm{II}(d)$ apart from the marginals are preserved. 

Now the number of terms in $\mathrm{LI}(d)$ with respect to the marginals can be calculated by simplifying the coefficients~\cite{ip2003some}:
\begin{equation*}
    (-1)^{d-1}(d-1)+\sum_{j=2}^{d-1}(-1)^{d-j}\binom{d-1}{j}=(-1)^{d-1}
\end{equation*}
This is the coefficient of the marginals in $II(D)$, therefore $\mathrm{LI}(d) = \mathrm{II}(d)$.

Next, we prove the second equivalence ($\mathrm{SI}(d) = \mathrm{LI}(d)$) in three different ways.
\paragraph{Information geometry } Given a factorisation of $d$ variables that consists of only two non-singleton blocks $\mathbb{P}_{\pi_s}=\mathbb{P}_{S_1}\mathbb{P}_{S_2}$, the KL divergence between itself and $\prod_{i=1}^d\mathbb{P}_{\pi_i}$ can be simplified:
\begin{align*}
    D_{\KL}\left(\mathbb{P}_{\pi_s}\|\prod_{i=1}^d\mathbb{P}_{\pi_i}\right) = D_{\KL}\left(\mathbb{P}_{S_1}\mathbb{P}_{S_2}\|\prod_{i=1}^d\mathbb{P}_{\pi_i}\right) = D_{\KL}\left(\mathbb{P}_{S_1}\|\prod_{s\in S_1}\mathbb{P}_{\pi_s}\right) + D_{\KL}\left(\mathbb{P}_{S_2}\|\prod_{s\in S_2}\mathbb{P}_{\pi_s}\right)
\end{align*}
E.g. let $\mathbb{P}_{\pi_s}=\mathbb{P}_{12}\mathbb{P}_{34}$,
\begin{align*}
    & D_{\KL}(\mathbb{P}_{12}\mathbb{P}_{34}\|\mathbb{P}_1\mathbb{P}_2\mathbb{P}_3\mathbb{P}_4)\\
    =&D_{\KL}(\mathbb{P}_{12}\mathbb{P}_{34}\| \mathbb{P}_{12}\mathbb{P}_3\mathbb{P}_4)+D_{\KL}(\mathbb{P}_{12}\mathbb{P}_3\mathbb{P}_4\| \mathbb{P}_1\mathbb{P}_2\mathbb{P}_3\mathbb{P}_4)\\
    =& D_{\KL}(\mathbb{P}_{12}\mathbb{P}_{34}\| \mathbb{P}_1\mathbb{P}_2\mathbb{P}_{34})+D_{\KL}(\mathbb{P}_1\mathbb{P}_2\mathbb{P}_{34}\| \mathbb{P}_1\mathbb{P}_2\mathbb{P}_3\mathbb{P}_4) \\
    =& D_{\KL}(\mathbb{P}_{34}\| \mathbb{P}_3\mathbb{P}_4)+D_{\KL}(\mathbb{P}_{12}\| \mathbb{P}_1\mathbb{P}_2)
\end{align*}

The decomposition is shown in Figure~\ref{fig: pythagorean}.
\begin{figure}[ht!]
    \centering
    \includegraphics[width=0.35\textwidth]{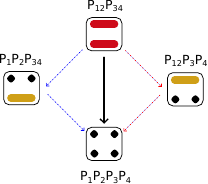}
    \caption{\textbf{Pythagorean theorem} KL divergence related to the factorisation of 4 variables is reduced to the sum of KL divergence related to 2 variables.  }
    \label{fig: pythagorean}
\end{figure}

If the number of non-singleton blocks is more than two, one can decompose two blocks at each time and get:
\begin{equation*}
    D_{\KL}\left(\prod_{j}\mathbb{P}_{S_j}\|\prod_{i=1}^d\mathbb{P}_{\pi_i}\right) = \sum_j D_{\KL}\left(\mathbb{P}_{S_j}\|\prod_{s\in S_j}\mathbb{P}_{s}\right)
\end{equation*}
where $\sum_j |S_j| = d$.
This is known as the Pythagorean theorem of KL divergence in information geometry~\cite{amari2001information}.
% \MB{I think this is only true for particular probability functions. It is for KL and linear families, I think. Otherwise, it is an inequality, no? Maybe I am mis-remembering}
% \MB{Maybe this is true because all the pdfs we are considering are part of the lattice, i.e., refinements/mergers of each other, and that makes it true?}
One can compute the coefficients of the resulting lower-order KL divergence terms and will see cancellations occur with the original lower-order terms. Instead here we show this can be easily achieved by the information measures defined on the interval lattices.
\paragraph{Interval lattice}
The recursiveness of $\mathrm{SI}(d)$ and $\mathrm{LI}(d)$ in Lemma~\ref{lemma: recurisve} shows that both measures can be expressed by a sum of themselves in lower-order. If we take the difference between $\mathrm{SI}(d)$ and $\mathrm{LI}(d)$ we get:
\begin{equation*}
    \mathrm{SI}(d) - \mathrm{LI}(d) = \sum_s \mathrm{SI}(\pi_s)
\end{equation*}
where $\mathrm{SI}(\pi_s)$ is the Streitberg information defined on the interval lattice $[\hat{\mathbf{0}}, \pi_s]$, a sublattice of $P(d)$ (for details of the measure on interval lattice, see Section~\ref{SI: generalised_interaction}). Each $\mathrm{SI}(\pi_s)$ is expressed by the sum of the KL divergences related to the factorisation of $\pi_s$ and equals zero due to the Pythagorean theorem above.
\paragraph{Expectation}
Finally, we define $\Delta^{B(d)}\mathbb{P}$ as the M\"obius inversion on $B(d)$ with the probability distribution function. Then we take the log of each term in $\Delta^{B(d)}\mathbb{P}$ and denote this as $\Delta^{B(d)}\log\mathbb{P}$. We note that $\mathrm{II}(d)$ can be constructed by taking the expectation of $\Delta^{B(d)}\log\mathbb{P}$. By the equivalence proved in Ref.~\cite{ip2003some}, we also get equivalence.
\end{proof}

\subsection{Proof of Lemma~\ref{lemma: recurisve}}\label{pf: lemma2}
\begin{proof}
    The Zeta function of $\mathrm{SI}(d)$ is:
    \begin{equation*}
        D_{\alpha}(\mathbb{P}_{1\ldots d}\|\prod_{i=1}^d \mathbb{P}_i) = \sum_{\pi} \mathrm{SI}(\pi)
    \end{equation*}
    for all $\pi\in P(d)$. Moving the lower-order sum to the left hand side we recover:
    \begin{equation*}
        \mathrm{SI}(d) = D_{\alpha}(\mathbb{P}_{1\ldots d}\|\prod_{i=1}^d \mathbb{P}_i)-\sum_{\pi\setminus \hat{\mathbf{1}}} \mathrm{SI}(\pi)
    \end{equation*}
    Similarly for $\mathrm{LI}(d)$.
\end{proof}

\section{Simplicial Complex}\label{sec: simplicial_complex}
We first show how simplicial complex is linked to the Boolean lattice.
The definition of a $k$-simplex $\theta$ is a set of $k$+1 vertices $\theta = [p_0, ..., p_k]$, i.e., a 1-simplex is a line, a 2-simplex is a triangle, 3-simplex is a tetrahedron, etc. A simplicial complex $\Theta$ is a collection of simplices that satisfy two conditions: 
(i) if $\theta \in \Theta$, then all the sub-simplices $v\subset \theta$ built from subsets of $\theta$ are also contained in $\Theta$; and (ii) the non-empty intersection of two simplices $\theta _{1},\theta _{2} \in \Theta$ is a sub-simplex of both $\theta _{1}$ and $\theta _{2}$. 
One easily realise that the simplicial complex with inclusion ordering is isomorphic to the Boolean lattice.

\begin{figure}[ht!]
    \centering
    \includegraphics[width=0.8\textwidth]{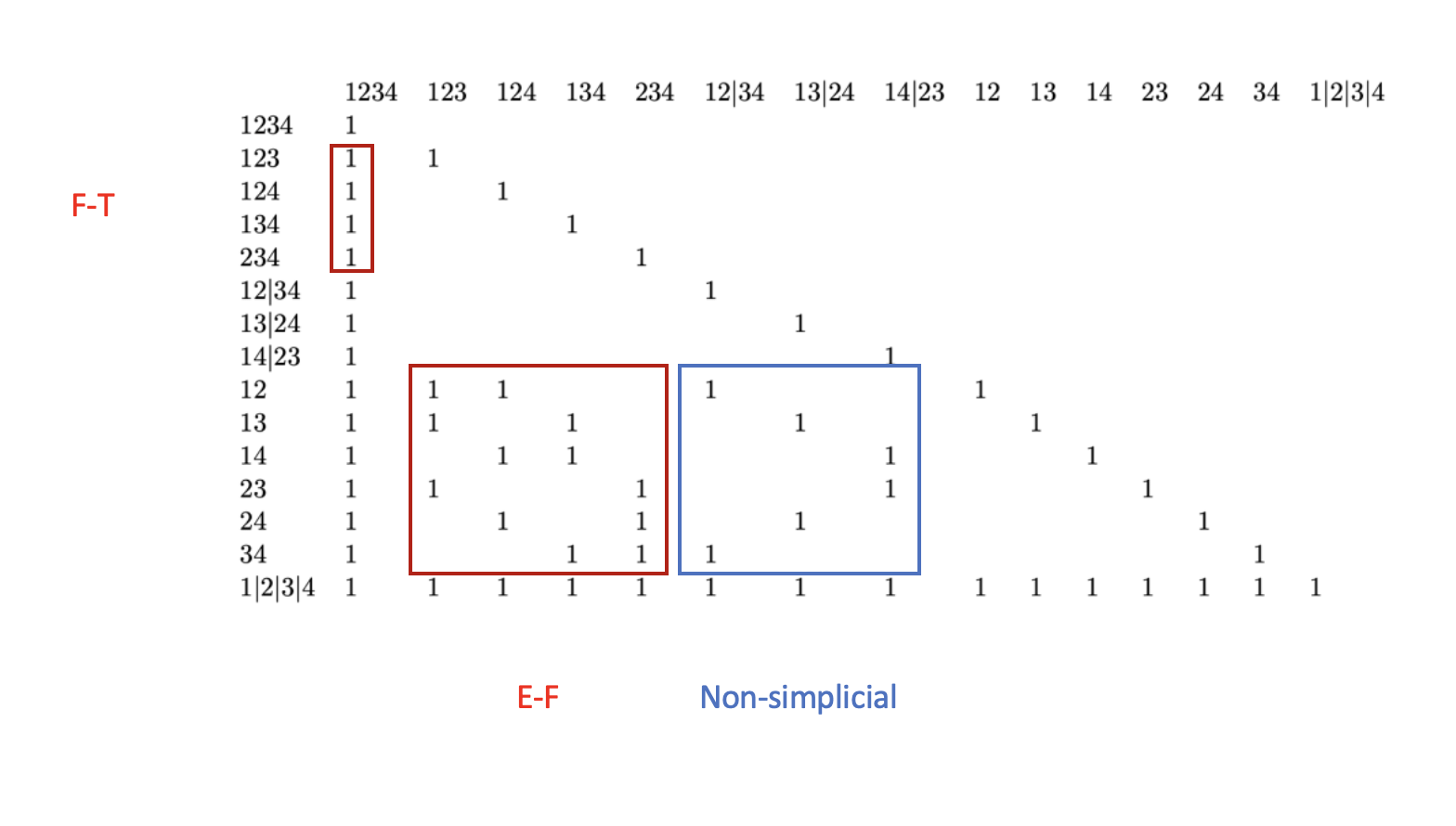}
    \caption{\textbf{Zeta matrix for the partition lattice of 4 variables.} Singletons in the partial factorisations are omitted for simplicity. The matrix can be decomposed into small blocks which correspond to the boundary operators in 3-simplex. Here the two red blocks represent the \textbf{E}dge-to-\textbf{F}ace operator and \textbf{F}ace-to-\textbf{T}etrahedron operator. The blue block corresponds to the non-simplicials, partitions with no singletons. }
    \label{fig:incidence}
\end{figure}

\begin{figure}[ht!]
    \centering
    \includegraphics[width=0.8\textwidth]{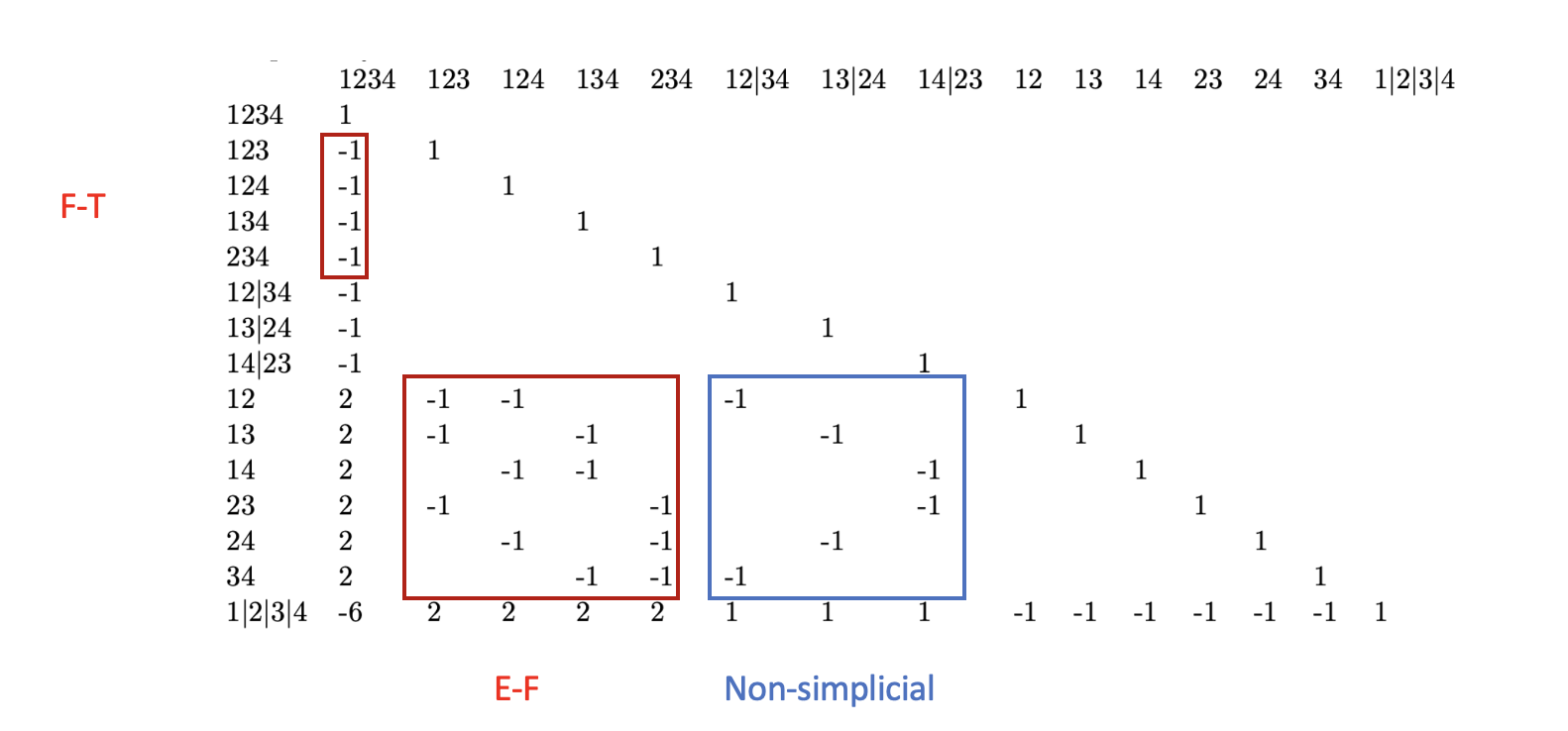}
    \caption{\textbf{M\"obius matrix for the partition lattice of 4 variables.} Comparing with the Zeta matrix above we see that the block structure is preserved.}
    % The value is completely modular meaning that it will be the same if the elements are in another poset, e.g. nonsimplicial part in the join-semilattice of the non-singletons.}
    \label{fig:mobius}
\end{figure}

Next we link the simplicial complex to the partition lattice using its boundary operators.
Below we have the Zeta matrix in Figure~\ref{fig:incidence} and M\"obius matrix in Figure~\ref{fig:mobius} to illustrate that the boundary operators (with no directionality) of ($d$-1)-simplex can be founded as block matrices within them. Simplicial complex does not contain terms consisting of more than one non-singleton block, e.g. $\mathbb{P}_{12}\mathbb{P}_{34}$
which are crucial in the partition lattice to ensure vanishment when $\mathbb{P}_{1234} = \mathbb{P}_{12}\mathbb{P}_{34}$.

\section{Generalised Streitberg information}\label{SI: generalised_interaction}
The Streitberg information defined on a factorsation $\mathbb{P}_{\pi}$ with $\pi\in P(d)$, $\mathrm{SI}(\pi)$ can be used to assess the factorisability of $\mathbb{P}_{\pi}$. We can formulate the generalised Streitberg information using the generalised Streitberg interaction measure proposed in Ref.~\cite{liu2023interaction}. Again we first find the M\"obius inversion on the interval lattice~\cite{mccullagh2018tensor, gratzer2011lattice} $[\hat{\mathbf{0}}, \pi]$ and apply the divergence function $D(\cdot\|\prod_{i=1}^{d}\mathbb{P}_i)$ to each term.
\begin{definition}[Generalised Streitberg information]
    \begin{equation*}
        \mathrm{SI}(\pi) = \sum_{\sigma\leq\pi}m(\sigma,\pi)D(\mathbb{P_{\sigma}}\|\prod_{i=1}^{d}\mathbb{P}_i)
    \end{equation*}
    where $\pi\in P(d)$ , $m$ are the M\"obius coefficients on $[\hat{\mathbf{0}}, \pi]$.
\end{definition}
This provides a general framework to detect higher-order information within blocks of variables. In fact, the generalised Streitberg information can be expressed in terms of ordinary Streitberg information in Definition~\ref{def: Li_SI}~\cite{mccullagh2018tensor}:
\begin{equation*}
\mathrm{SI}(\pi)=\sum_{\sigma \vee \pi=\hat{\mathbf{1}}} \prod_{b\in\sigma}\mathrm{SI}(\sigma).
\end{equation*}

\section{Additional synthetic experiments}\label{SI: order35_experiments}

\begin{figure}[ht!]
    \centering
    \includegraphics[width=0.95\textwidth]{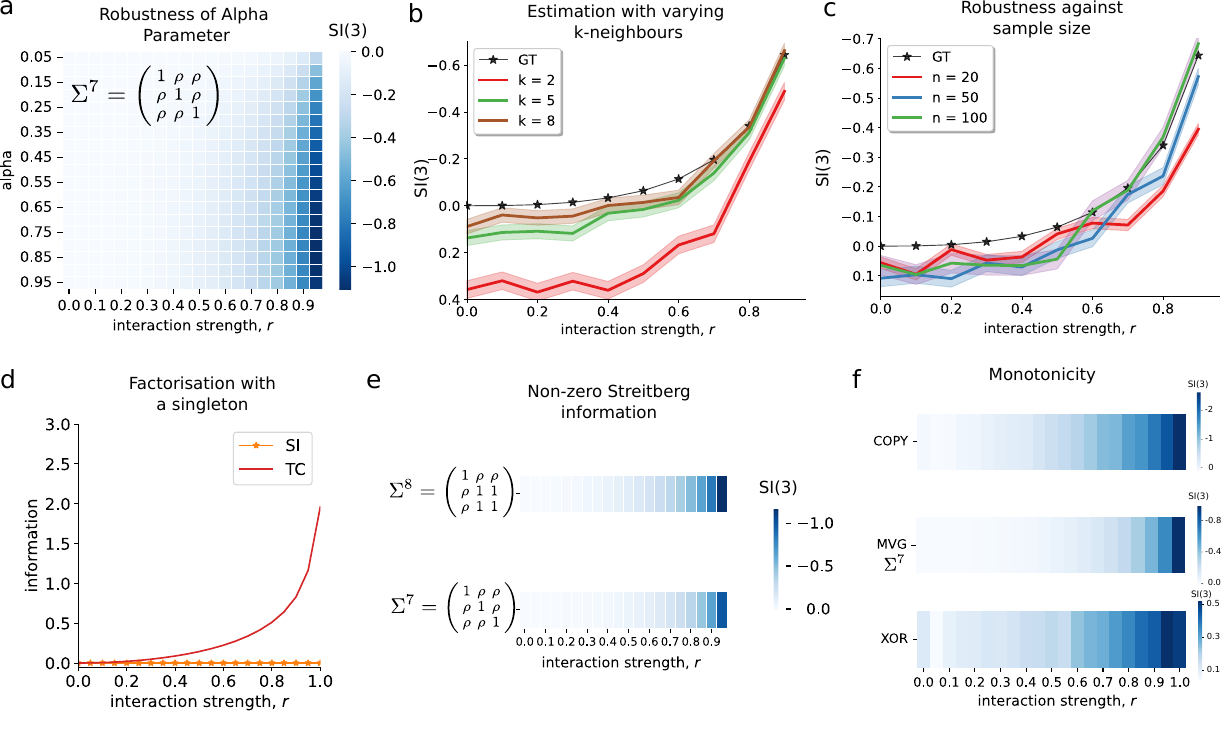}
    \caption{
    \textbf{Synthetic experiments of 3-order Streitberg information}
     \textbf{a} The behaviour of $\mathrm{SI}(3)$ is consistent across choice of $\alpha$. 
    \textbf{b} Increasing the number of neighbours $k$ improve estimation accuracy. 
    \textbf{c} Increasing the sample size $n$ improves estimation accuracy.
     \textbf{d} $\mathrm{TC}(3)$ fails to vanish for $\mathbb{P}_{1}\mathbb{P}_{23}$.
    \textbf{e} The magnitude of $\mathrm{SI}(3)$ is influenced by the extent to which the joint distribution can be factorised.
    \textbf{f} Streitberg Information exhibits monotonic behaviour which is consistent across varying types of interaction.}
    \label{fig: order_3}
\end{figure}

\begin{figure}[ht!]
    \centering
    \includegraphics[width=0.95\textwidth]{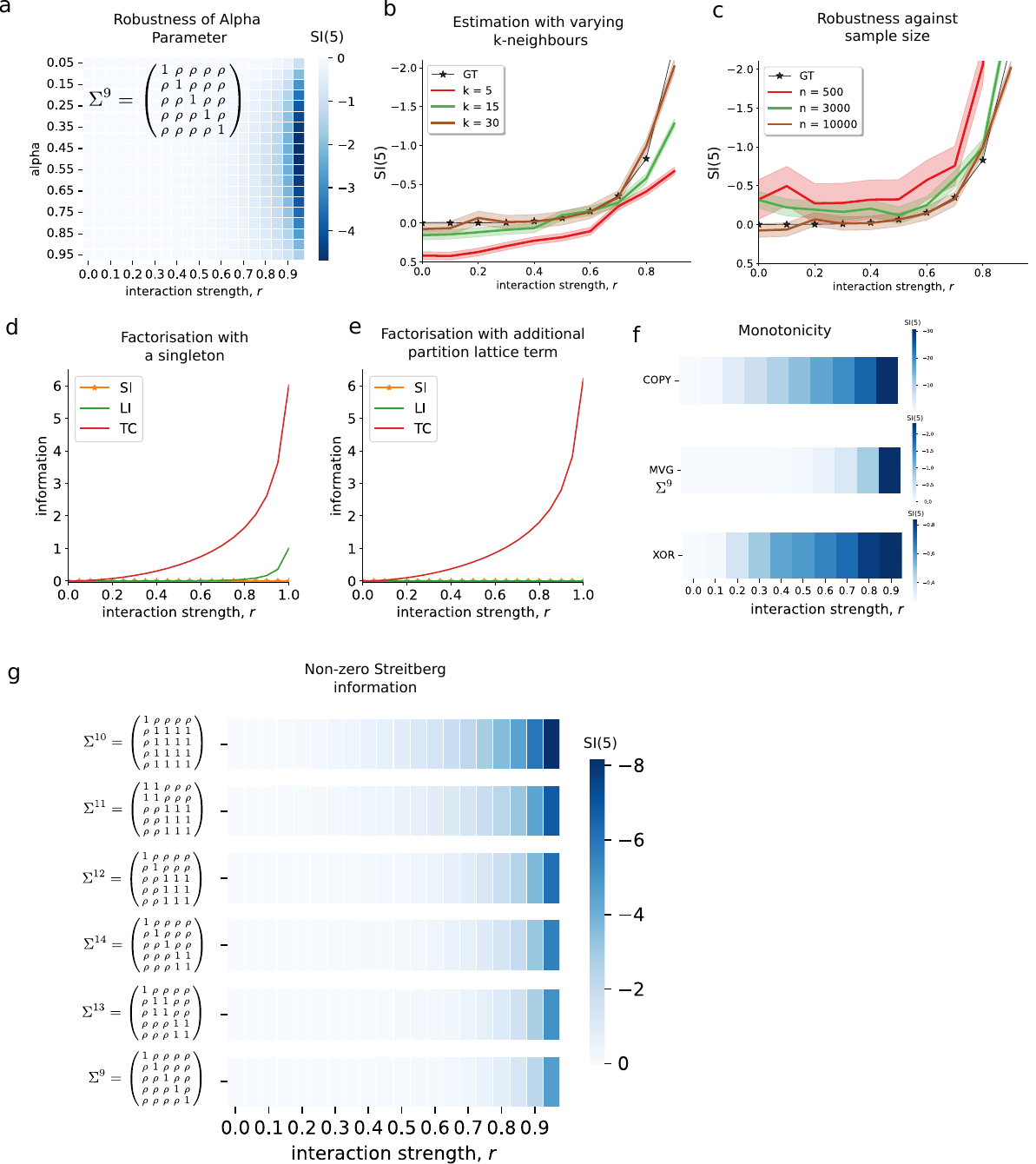}
    \caption{\textbf{Synthetic experiments of 5-order Streitberg information}
     \textbf{a} The behaviour of $\mathrm{SI}(5)$ as a function of interaction strength is consistent across choice of $\alpha$. 
    \textbf{b} Increasing the number of neighbours $k$ improve estimation accuracy. 
    \textbf{c} Increasing the sample size $n$ improves estimation accuracy.
     \textbf{d} $\mathrm{TC}(5)$ fails to vanish for $\mathbb{P}_{1}\mathbb{P}_{2345}$.
    \textbf{e} both $\mathrm{TC}(5)$ and $\mathrm{LI}(5)$ fail to characterise $\mathbb{P}_{12}\mathbb{P}_{345}$. $\mathrm{SI}(5)$ correctly vanishes in both cases.
    \textbf{f} The magnitude of $\mathrm{SI}(5)$ is influenced by the extent to which the joint distribution can be factorised. 
    \textbf{g} Streitberg Information exhibits monotonic behaviour which is consistent across varying types of interaction.
   }
    \label{fig: order_5}
\end{figure}
\newpage
\section{Tsallis-Alpha Divergence for Multivariate Gaussian: Analytical Expression}\label{sec: tsallis}
Following from the analytical solution of Tsallis-Alpha divergence between multivariate Gaussian~\cite{poczos2011estimation}, we further simplify the divergence between the multivariate Gaussian with $\Sigma_f$ and its complete factorisation with $\Sigma_g = I_d$.
\begin{align*}
    D_\alpha\left(\mathbb{P}_{1\cdots d} \| \prod_{i=1}^d \mathbb{P}_i\right)=&\frac{1}{\alpha - 1}\left(\int f^\alpha(x) g^{1-\alpha}(x) \mathrm{d} x - 1\right)\\
    =&\frac{\left|{\Sigma}_f^{-1}\right|^{\alpha / 2}\left|{\Sigma}_g^{-1}\right|^{(1-\alpha) / 2}}{\left|\alpha {\Sigma}_f^{-1}+(1-\alpha) {\Sigma}_g^{-1}\right|^{1 / 2}} = 
    \frac{\left|{\Sigma}_f^{-1}\right|^{\alpha / 2}\left|I_d\right|^{(1-\alpha) / 2}}{\left|\alpha {\Sigma}_f^{-1}+(1-\alpha) I_d\right|^{1 / 2}}
    = \frac{\left|{\Sigma}_f^{-1}\right|^{\alpha / 2}}{\left|\alpha {\Sigma}_f^{-1}+(1-\alpha) I_d\right|^{1 / 2}}
\end{align*}
where $|\cdot|$ is the determinant of the matrix. Here $\Sigma_g$ is simply the identity matrix in our case, i.e. mean vector is the vector of zero and the covariance matrix has no pairwise dependence and the variances on the diagonal are 1. 

Specifically, when $\alpha=1$, we retrieve the KL divergence and its analytical solution of multivariate Gaussian is
\begin{align*}
    D_{\KL}\left(\mathbb{P}_{1\cdots d} \| \prod \mathbb{P}_i\right) = \frac{1}{2}\left( -\log {\left|\boldsymbol{\Sigma}_g\right|}+ \operatorname{tr}\left[\boldsymbol{\Sigma}_f \right]-d\right)
    =-\frac{1}{2}\log {\left|\boldsymbol{\Sigma}_g\right|}
\end{align*}

\section{Stock market data}\label{sec: stockmarket}
The data is accessed using the following link: \url{https://www.kaggle.com/datasets/andrewmvd/sp-500-stocks}. 
\begin{figure}[ht!]
    \centering
    \includegraphics[width=0.95\textwidth]{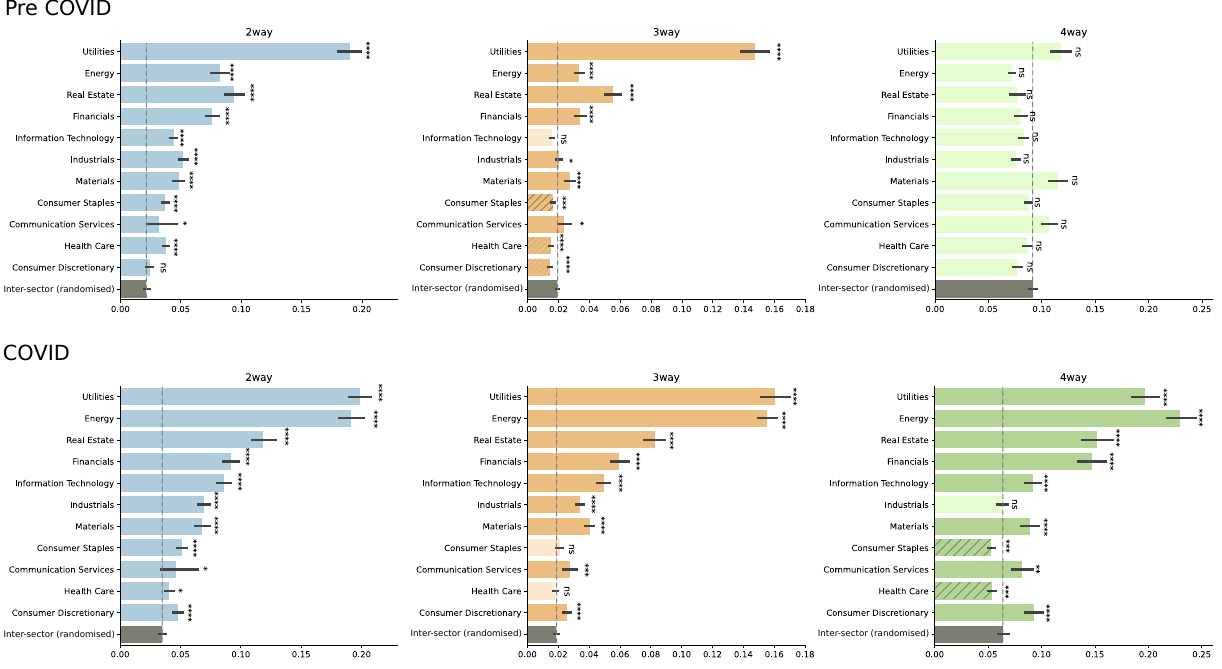}
    \caption{
    \textbf{$d$-order Streitberg information of stock returns before and after COVID.}
    All the cleaned data before/after 2020 are used to compute the higher-order information between stocks for the pre-COVID/COVID period.
    }
    \label{fig: sp500_sup}
\end{figure}
For a given stock in the S\&P 500, we calculate the daily return from 4 Jan 2010 to 24 Apr 2024 as the difference between its closing and opening prices on each day. The returns at different time points are generally assumed as $iid$~\cite{ali1982identical}.
Next we drop the stocks that have more than 20\% of missingness in time and then drop the time points if there is missingness in any stocks. As a result we have 2900 realisations of daily returns of 459 stocks in S\&P 500. Each stock belongs to one of the 11 sectors as defined in Global Industry Classification Standard (GICS). We compute the pairwise, 3-, 4-way $\mathrm{SI}$ to the sets of stocks that are either taken from within the same sector or sampled at random from the 459 stocks. In each case, we sample 500 sets of regions or take all possible combinations, whichever is lowest. 

\newpage
\section{Macaque reaching task data}\label{SI:monkey}
\begin{figure}[ht!]
    \centering
    \includegraphics[width=0.25\textwidth]{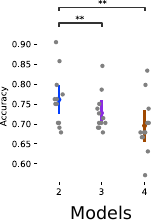}
    \caption{
    \textbf{Accuracy of macaque decoding models using $d$-order interactions separately.}
    Increasing order interactions contained less information for predicting preparation vs action stages.   }
    \label{fig: monkey_SI}
\end{figure}
In the model predicting the preparation and action stages, we use set the time points that correspond to preparation and action to 0 and 1 respectively. In the model decoding the direction of the monkey's hand movement, we use use ordinal encoding to label the directions from DownLeft to DownRight using the order in Figure~\ref{fig: monkey}a. We then train each Logistic classifier (with liblinear solver and L2 penalty) using the 80\% of the data and report the model accuracy on the other 20\%. The data can be accessed through \url{https://dataverse.harvard.edu/file.xhtml?fileId=6969883&version=11.0}.

% \section{Feature selection}

\section{PID embedding}\label{SI: PID}
\begin{figure}[ht!]
    \centering
    \includegraphics[width=0.95\textwidth]{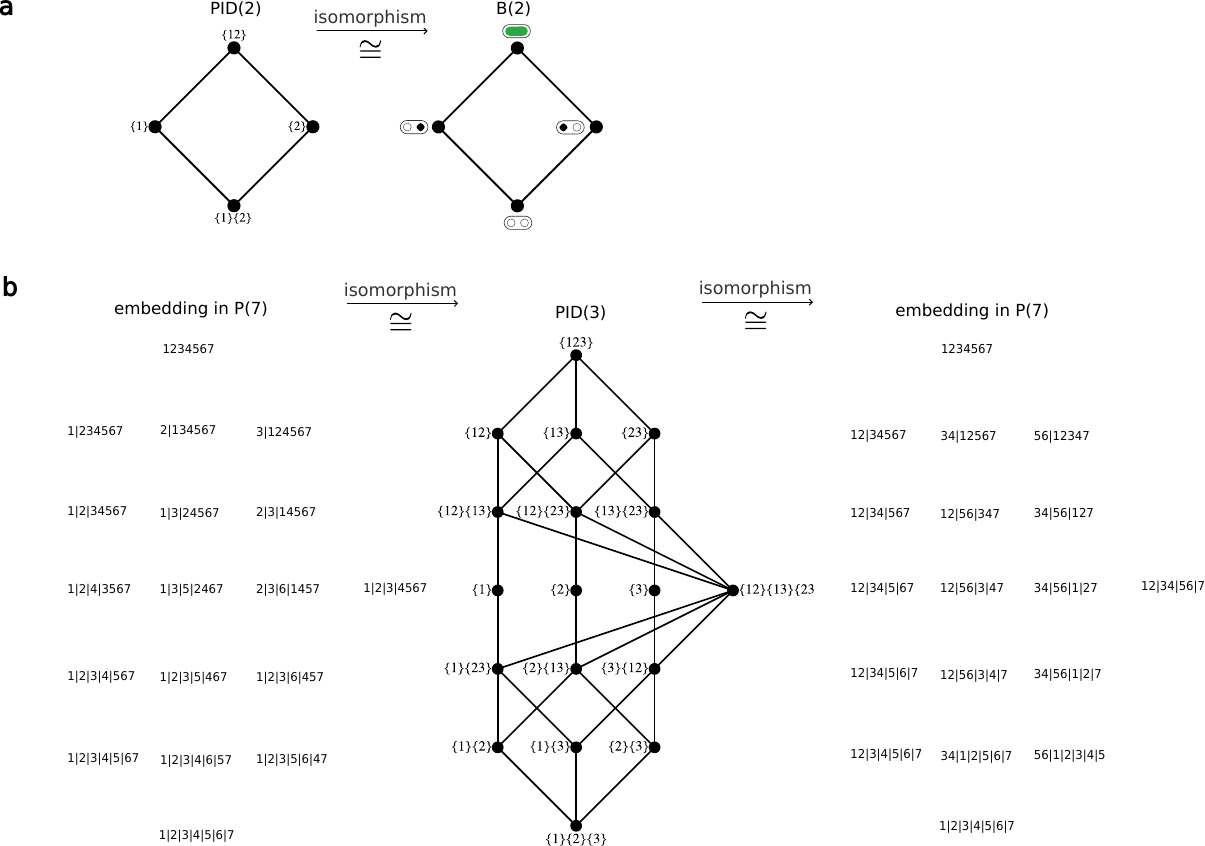}
    \caption{\textbf{Isomorphisms of free distributive lattice.} \textbf{a} The PID lattice of two variables is isomorphic to the $B(2)$. \textbf{b} The PID lattice of three variables can be embedded in $P(7)$. Two embeddings are shown.}
    \label{fig: PID}
\end{figure}
Multivariate information measures are frequently used to measure synergy and redundancy, however, the precise meanings of synergy and redundancy is an ongoing debate.
A typical example of synergy is an XOR logic gate whilst a typical example of redundancy is when all the variables are identical. 
PID offers a direct approach to quantify synergy and redundancy, defining them as distinct information atoms in the mutual information between a target variable and a set of source variables~\cite{williams2010nonnegative}.
The entire set of information atoms are formalised by the redundancy lattice (the free distributive lattice without its maximal and minimum elements), with the number of information atoms given by the Dedekind number, which quickly becomes intractable (e.g., 7581 when $d=5$; the exact values beyond $d=9$ has yet to be found~\cite{jakel2023computation}).
To the best of our knowledge, due to the complexity of the free distributive lattice, PID is not currently defined for $d>3$ and will be computationally prohibitive due to the lattice structure and the directed construction.

Notably, it has been shown that every finite lattice can be embedded into a finite partition lattice~\cite{pudlak1980every}, leading to the following relationship between the free distributive lattice and the partition lattice:
\begin{lemma}
    A free distributive lattice with length $d$ can be embedded into a partition lattice of length $d$ where the length of a lattice $L$ is $d$ if there is a chain in $L$ of length $d$ and all chains in $L$ are of length $\leq d$. 
\end{lemma}

This is a simple result from the tight embedding property of distributive lattices in the partition lattice~\cite{wild2018tight}. In particular, the PID lattice with two variables, excluding empty joins and meets, is isomorphic to the Boolean lattice with two elements (in Figure~\ref{fig: PID}a). Similarly, the PID lattice with three variables can be embedded into a partition lattice with seven elements (two examples shown in Figure~\ref{fig: PID}b).

\newpage
\section{Computational considerations}\label{SI: compute}
The $k$NN-based estimation of the $D_{\alpha}$ has time complexity $\mathcal{O}(dkn)$ where $d$ is the number of variables, $k$ is the parameter in $k$NN and $n$ is the sample size. 
The time complexity for estimating $D_{\alpha}$ can be reduced as the singletons in each partition cancel out with the singletons in the complete factorisation. E.g. when $d=3$, 
\begin{align}
    \mathrm{SI}(3) =& \mathrm{LI}(3)\\
    =& D_{\alpha}(\mathbb{P}_{123}\|\mathbb{P}_1\mathbb{P}_2\mathbb{P}_3) - D_{\alpha}(\mathbb{P}_1\mathbb{P}_{23}\|\mathbb{P}_1\mathbb{P}_2\mathbb{P}_3) - D_{\alpha}(\mathbb{P}_2\mathbb{P}_{13}\|\mathbb{P}_1\mathbb{P}_2\mathbb{P}_3) \\
    &- D_{\alpha}(\mathbb{P}_3\mathbb{P}_{12}\|\mathbb{P}_1\mathbb{P}_2\mathbb{P}_3) + 2 D_{\alpha}(\mathbb{P}_1\mathbb{P}_2\mathbb{P}_3\|\mathbb{P}_1\mathbb{P}_2\mathbb{P}_3)\label{eqn: si3}\\
    =& D_{\alpha}(\mathbb{P}_{123}\|\mathbb{P}_1\mathbb{P}_2\mathbb{P}_3) - D_{\alpha}(\mathbb{P}_{23}\|\mathbb{P}_2\mathbb{P}_3) - D_{\alpha}(\mathbb{P}_{13}\|\mathbb{P}_1\mathbb{P}_3) - D_{\alpha}(\mathbb{P}_{12}\|\mathbb{P}_1\mathbb{P}_2)\label{eqn: si3_simplify}
\end{align}
The cancellation of singletons in Equation~\eqref{eqn: si3_simplify} reduces the complexity from estimating five 3-dimensional $D_{\alpha}$  (in total 15) to one 3-dimensional and three 2-dimensional $D_{\alpha}$ (in total 9). Here we fix $k$ and $n$ and investigate how singleton cancellations reduce the time complexity of $\mathrm{SI}(d)$ and $\mathrm{LI}(d)$ as $d$ varies between 2 and 6 shown in Table~\ref{tab: time_complexity}.
\begin{table}[ht!]
\centering
\caption{Computational improvement after cancellations of singletons.}
\label{tab: time_complexity}
\begin{tabular}{llllll}
\hline
Order & 2   & 3    & 4     & 5       & 6         \\ \hline
$\mathrm{LI}(d)\Rightarrow(cancellation)$    & $2\Rightarrow(2)$ & $15\Rightarrow(9)$ & $48\Rightarrow(28)$ & $135\Rightarrow(75)$  & $348\Rightarrow(186) $  \\ 
$\mathrm{SI}(d)\Rightarrow(cancellation)$    & $2\Rightarrow(2)$ & $15\Rightarrow(9)$ & $60\Rightarrow(40)$ & $260\Rightarrow(210)$ & $1236\Rightarrow(1056)$ \\ \hline
\end{tabular}
\end{table}

We report the computational time in Table~\ref{tab: time_taken} for all experiments of Streitberg information using the nonparametric estimation . All experiments carried out on a 2015 iMac with 4 GHz Quad-Core Intel Core i7 processor and 32 GB 1867 MHz DDR3 memory.
\begin{table}[ht!]
\centering
\caption{Time taken for experiments.}
\label{tab: time_taken}
\begin{tabular}{lllllll}
\hline
Experiment & Fig.2b & Fig.2c & Fig.3d & Fig.4a  & Fig.4b & Fig.5    \\ \hline
T            ime       & 22m35s & 1m28s  & 9m54s  & 1h26m3s & 50m2s  & 5h19m32s \\ \hline
\end{tabular}
\end{table}
The code and data used to perform the experiments in the paper are provided in the GitHub repository:
\url{https://github.com/barahona-research-group/streitberg-information.git}.

\end{document}